\documentclass[prd,aps,twocolumn,showpacs,preprintnumbers,nofootinbib]{revtex4}
\usepackage{eurosym}
\usepackage[dvips]{graphicx}
\usepackage{epsf}
\usepackage{amsmath}
\usepackage{amssymb}
\usepackage{graphicx}
\usepackage{dcolumn}
\usepackage{bm}

\begin{document}

\title{The $AdS_{5}\times S^{5}$ fermionic model}
\author{Elcio Abdalla}
\affiliation{Instituto de Fisica, Universidade de Sao Paulo, CP 66318, 05315-970, Sao
Paulo, Brazil}
\author{Antonio Lima-Santos}
\affiliation{Dept. de Fisica, Univ. Fed. Sao Carlos, CP 676, CEP 13569-905, Sao Carlos,
Brazil }

\begin{abstract}
We consider the $AdS_{5}\times S^{5}$ integrable model. As it turns out,
relying on well known arguments, we claim that the conformally invariant
fermionic model is solvable, the resulting solution given in terms of two
current algebras realizations.
\end{abstract}

\maketitle

\section{Introduction}

Integrable models have a long and successful history \cite{aar}. In
particular, models defined on a symmetric space are generally integrable 
\cite{eichforger,abdetal,afg}. This means that an infinite number of local
conservation laws exist \cite{eichforger}, or at least one nonlocal
conservation law \cite{aar,luscherpohl}. In general, such integrable models
display a non vanishing mass gap, useful for describing the exact S-matrix
in terms of rapidities \cite{zamzam,abdetal}. In such a line, a large number
of models have been solved and their exact on shell solution obtained \cite%
{eals,lsmm}.

There is also at least one model where no mass gap exists, but comprising
non trivial conservation laws. It is the case of the chiral Gross-Neveu
model \cite{cgn}. Supposing the existence of a mass gap, the model has been
solved on shell \cite{abwkks}. However, it is known that there is a non
trivial fix point such that the theory allows for a conformally invariant
solution as well, for a given value of the coupling constant \cite{df}.

This means that an integrable model can also contain a conformally invariant
solution. This is a quite non trivial fact that we wish to explore in case
of integrable models relevant for string theory, where conformal invariance
is a very desirable property.

In the framework of string theory, it is possible to gather information
about the Yang-Mills theory at intermediate coupling. Obtaining a strongly
coupled field theory underlying the {\small QCD} string actually provides an
integrable model in the world sheet, and the low dimensionality of the
problem may imply exact solvability \cite{bena}.

In that case, the symmetry of the integrable model is $\frac{PSU(2,2|4)}{%
SO(4,1)\times SO(5)}$. The bosonic part of such a coset is $AdS_{5}\times
S^{5}$, which will be our main concern. Most of the literature is related,
in this case, to integrable models and their nonlocal conservation laws \cite%
{nonlocalads5}. Currents for the pure spinor superstring in $AdS_{5}\times
S^{5}$ have subsequently been constructed \cite{val}. While the role of $%
AdS_{5}$ is largely discussed in relation to string solutions \cite%
{mandaletal,bmn}, integrable structures are related to the underlying string
spectrum \cite{hatsudayoshida}.

Later, the non local charges have been also been related to a {\small BRST}
cohomology \cite{klusonberko} ensuring $\kappa $-symmetry. One thus
conjectured that conformal invariance should be related to the integrable
models relevant to string theory.

On the other hand, in string theory, a lot has been done concerning
integrability of the underlying symmetry of strings in certain backgrounds.
In Maldacena's conjecture, four dimensional $N=4$ super Yang-Mills theory is
dual to super strings in $AdS_{5}\times S^{5}$ background \cite{malda97}.
But 
\begin{equation}
AdS_{5}\times S^{5}\equiv \frac{SO(5,1)}{O(4,1)}\times \frac{SO(6)}{O(5)}.
\label{ads}
\end{equation}%
This means that the model is defined on a symmetric space, thus implying a
non trivial (and non local) conservation law \cite{eichforger}. Moreover,
since the symmetric space is a direct product of symmetric spaces with
simple gauge groups, the sigma model defined on that space is also
integrable at the quantum level \cite{afg}. On the other hand, conformal
invariance is very useful in string theory and the question is whether these
models display conformal invariance, at least in some form. The answer is
positive, as we show.

We shall consider a fermionic model defined upon the space (\ref{ads}).
Following old and well established arguments we see that at a well defined
value of the coupling constant the theory is conformally invariant.

\section{Conserved currents}

The above mentioned fermionic model is defined by the lagrangian density%
\begin{equation}
L=\overset{\_}{\psi }_{ia}i\gamma ^{\mu }\partial _{\mu }\psi
_{ia}+g_{1}J_{\mu ab}J_{ba}^{\mu }+g_{2}J_{\mu ij}J_{ji}^{\mu }
\end{equation}%
where we define the currents are given by $J_{\mu \ ab}=\overset{\_}{\psi }%
_{ia}\gamma _{\mu }\psi _{ib}$ and $J_{\mu \ ij}=\overset{\_}{\psi }%
_{ia}\gamma _{\mu }\psi _{ja}$. They are related to the first or second
factors defining the underlying symmetry group, that is, we identify the
labels $a,b,...$ as being in $SO(5,1)$ and $i,j,...$ in $SO(6)$. Here, $%
g_{1} $ and $g_{2}$ are, up to now, arbitrary coupling constants.

The field equation for $\psi _{ia}$ is 
\begin{equation}
i\gamma ^{\mu }\partial _{\mu }\psi _{ia}=-2g_{1}\gamma ^{\mu }\psi
_{ic}J_{\mu \ ca}-2g_{2}\gamma ^{\mu }\psi _{ka}J_{\mu \ ki}\quad ,
\end{equation}%
while 
\begin{equation}
i\partial _{\mu }\overset{\_}{\psi }_{ia}\gamma ^{\mu }=2g_{1}J_{\mu \ ac}%
\overset{\_}{\psi }_{ic}\gamma ^{\mu }+2g_{2}J_{\mu \ ik}\overset{\_}{\psi }%
_{ka}\gamma ^{\mu }
\end{equation}%
is the field equation for $\overset{\_}{\psi }_{ia}$.

The Noether currents related to the symmetries $SO(5,1)$ and $SO(6)$,
respectively, obey the conservation equations 
\begin{equation}
\partial _{\mu }J_{\ ab}^{\mu }=0\quad \mathrm{and}\quad \ \partial _{\mu
}J_{\ ij}^{\mu }=0\quad .
\end{equation}

Let us now consider the axial currents (non)conservation laws. Using the
relations for the $\gamma _{\mu }$ matrices we have 
\begin{eqnarray}
\epsilon ^{\mu \nu }\gamma _{\nu } &=&\gamma ^{5}\gamma ^{\mu }\; ,\quad
\epsilon ^{01}=1 \quad ,  \notag \\
(\gamma ^{5})^{2} &=&1\; ,\quad (\gamma ^{0})^{2}=1\; ,\quad (\gamma
^{1})^{2}=-1 \quad ,  \notag \\
\gamma ^{1} &=&-\gamma _{1},\quad \gamma ^{5}=\gamma ^{0}\gamma ^{1},\quad
\gamma ^{5}\gamma ^{0}=\gamma _{1}\quad .
\end{eqnarray}%
We can compute the divergence of the axial current, $\epsilon ^{\mu \nu
}\partial_\mu J_{\nu \ ab}$, 
\begin{eqnarray}
&&\epsilon ^{\mu \nu }i\partial _{\mu }\left( \overset{\_}{\psi }_{ia}\gamma
_{\nu }\psi _{ib}\right)  \notag \\
&=&i\partial _{\mu }\left( \overset{\_}{\psi }_{ia}\gamma ^{5}\gamma ^{\mu
}\psi _{ib}\right)  \notag \\
&=&-[i\partial _{\mu }\overset{\_}{\psi }_{ia}\gamma ^{\mu }]\gamma ^{5}\psi
_{ib}+\overset{\_}{\psi }_{ia}\gamma ^{5}[i\partial _{\mu }\gamma ^{\mu
}\psi _{ib}].
\end{eqnarray}%
Taking into account the field equations we get%
\begin{eqnarray}
&&\epsilon ^{\mu \nu }i\partial _{\mu }\left( \overset{\_}{\psi }_{ia}\gamma
_{\nu }\psi _{ib}\right)  \notag \\
&=&-2g_{1}[\overset{\_}{\psi }_{ia}\gamma ^{5}\gamma ^{\mu }\psi
_{ic}]J_{\mu \ cb}-2g_{2}\overset{\_}{[\psi }_{ia}\gamma ^{5}\gamma ^{\mu
}\psi _{jb}]J_{\mu \ ji}  \notag \\
&&+2g_{1}J_{\mu \ ac}[\overset{\_}{\psi }_{ic}\gamma ^{\mu }\gamma ^{5}\psi
_{ib}]-2g_{2}J_{\mu \ ij}[\overset{\_}{\psi }_{ja}\gamma ^{\mu }\gamma
^{5}\psi _{ib}].  \notag \\
&&
\end{eqnarray}%
Here we note that the terms with the $g_{1}$ coefficient are products of two
currents while the terms with $g_{2}$ coefficient are cancelled, that is, 
\begin{eqnarray}
&&\epsilon ^{\mu \nu }i\partial _{\mu }\left( \overset{\_}{\psi }_{ia}\gamma
_{\nu }\psi _{ib}\right)  \notag \\
&=&-2g_{1}\epsilon ^{\mu \nu }[J_{\nu \ ac}J_{\mu \ cb}-J_{\mu \ ac}J_{\mu \
cb}]  \notag \\
&&-2g_{2}\epsilon ^{\mu \nu }[\overset{\_}{\psi }_{ia}\gamma _{\nu }\psi
_{jb}J_{\mu \ ji}-J_{\mu \ ij}\overset{\_}{\psi }_{ja}\gamma _{\nu }\psi
_{ib}]\quad .
\end{eqnarray}%
Using the identity 
\begin{equation}
\epsilon ^{\mu \nu }\left( \gamma _{\mu }\right) _{\alpha \beta }\left(
\gamma _{\nu }\right) _{\gamma \delta }=\delta _{\alpha \delta }\left(
\gamma _{5}\right) _{\beta \gamma }-\left( \gamma _{5}\right) _{\alpha
\delta }\delta _{\beta \gamma }\quad ,
\end{equation}%
the final result is%
\begin{equation}
\epsilon ^{\mu \nu }\partial _{\mu }\left( J_{\nu }\right)
_{ab}-4ig_{1}\epsilon ^{\mu \nu }\left( J_{\mu }J_{\nu }\right) _{ab}=0\quad
.
\end{equation}%
Therefore, the axial current $J_{\nu \ ab}^{(5)}=\epsilon ^{\mu \nu }J_{\nu
\ ab}$ fails to be conserved classically.

A similar result follows for the axial current $J_{\nu \ ij}^{(5)}=\epsilon
^{\mu \nu }J_{\nu \ ij}$, 
\begin{equation}
\epsilon ^{\mu \nu }\partial _{\mu }\left( J_{\nu }\right)
_{ij}-4ig_{2}\epsilon ^{\mu \nu }\left( J_{\mu }J_{\nu }\right) _{ij}=0\quad
.
\end{equation}

We consider now the axial anomaly contribution to the field equations. We
introduce the gauge field $\left( A_{\mu }\right) _{ab}=2g_{1}\left( J_{\mu
}\right) _{ab}$ in order to identify the anomaly term%
\begin{eqnarray}
\frac{N}{2\pi }\epsilon ^{\mu \nu }F_{\mu \nu } &=&\frac{N}{\pi }\epsilon
^{\mu \nu }\left( \partial _{\mu }A_{\nu }-iA_{\mu }A_{\nu }\right)  \notag
\\
&=&\frac{N}{\pi }[2g_{1}\epsilon ^{\mu \nu }\partial _{\mu }J_{\nu
}-i(2g_{1})^{2}J_{\mu }J_{\nu }\, ,
\end{eqnarray}%
to be added to the divergence equation for $J_{\nu \ ab}^{(5)}$, 
\begin{equation}
\epsilon ^{\mu \nu }\partial _{\mu }(J_{\nu })_{ab}-4ig_{1}\epsilon ^{\mu
\nu }(J_{\mu }J_{\nu })_{ab}=\frac{N}{2\pi }\epsilon ^{\mu \nu }(F_{\mu \nu
})_{ab}\; ,
\end{equation}%
where $N$ is the number of species, in this case equal to $6$. We are thus
led to 
\begin{equation}
\epsilon ^{\mu \nu }\partial _{\mu }(J_{\nu })_{ab} =4ig_{1}\frac{\left( 1-%
\frac{g_{1}N}{\pi }\right)}{\left( 1+\frac{2g_{1}N}{ \pi }\right) } \epsilon
^{\mu \nu }(J_{\mu }J_{\nu })_{ab}\; .
\end{equation}%
Therefore, the choice $g_{1}=\frac{\pi }{N}$ implies that the axial current
is also conserved, 
\begin{equation}
\epsilon ^{\mu \nu }\partial _{\mu }(J_{\nu })_{ab}=0 \quad .
\end{equation}%
This means conformal invariance in the coset $\frac{SO(5,1)}{O(4,1)}$.
Notice that, \textit{mutatis mutandis} we get similar a result for the $%
\frac{{SO(6)}}{{O(5)}}$ factor, as well as conformal invariance for all
spaces of the kind $AdS_{p}\times S^{q}$ in case we carefully choose the
coupling. Thus, at the point $g_{2}=\frac{2\pi }{6}$ the second axial
current is conserved%
\begin{equation}
\epsilon ^{\mu \nu }\partial _{\mu }(J_{\nu })_{ij}=0
\end{equation}%
and the fermionic theory in the coset $\frac{SO(6)}{O(5)}$ is conformally
invariant.

An alternative and equivalent proof of conformal invariance at a given
coupling can be obtained by arguments already known in \cite{df}. Thus, for
these values of $g_{1}$ and $g_{2}$ we get the conformal field $\psi _{ia}$
with $\frac{SO(5,1)}{SO(4,1)}\times \frac{SO(6)}{SO(5)}$ conformal
invariance.

\section{ Current Algebra}

We can write the equal-time commutation rules 
\begin{eqnarray}
\lbrack J_{0\ ab}(t,x),J_{0\ cd}(t,y)] &=&if_{abcd}^{ef}J_{0\ ef}(t,x)\delta
(x-y)  \notag \\
\lbrack J_{0\ ab}(t,x),J_{1\ cd}(t,y)] &=&if_{abcd}^{ef}J_{0\ ef}(t,x)\delta
(x-y)  \notag \\
&&+iC_{1}\delta _{ac}\delta _{bd}\delta ^{\prime }(x-y)  \notag \\
&&+iC_{2}\delta _{ad}\delta _{bc}\delta ^{\prime }(x-y)  \notag \\
\lbrack J_{1\ ab}(t,x),J_{1\ cd}(t,y)] &=&if_{abcd}^{ef}J_{0\ ef}(t,x)\delta
(x-y)  \notag \\
&&
\end{eqnarray}%
where $C_{1}$ and $C_{2}$ ($=0$ $\mathrm{or}-C_{1}$) are c-number Schwinger
terms. In addition, we also have%
\begin{eqnarray}
\lbrack J_{0\ ij}(t,x),J_{0\ kl}(t,y)] &=&if_{ijkl}^{pq}J_{0\ pq}(t,x)\delta
(x-y)  \notag \\
\lbrack J_{0\ ij}(t,x),J_{1\ kl}(t,y)] &=&if_{ijkl}^{pq}J_{0\ pq}(t,x)\delta
(x-y)  \notag \\
&&+iD_{1}\delta _{ik}\delta _{jl}\delta ^{\prime }(x-y)  \notag \\
&&+iD_{2}\delta _{il}\delta _{jk}\delta ^{\prime }(x-y)  \notag \\
\lbrack J_{1\ ij}(t,x),J_{1\ kl}(t,y)] &=&if_{ijkl}^{pq}J_{0\ pq}(t,x)\delta
(x-y)  \notag \\
&&
\end{eqnarray}%
where $D_{1}$ and $D_{2}$ ($=0$ $\mathrm{or}-D_{1}$) are also c-number
Schwinger terms.

Here we note the structure constants $f_{abcd}^{ef}$ of the factor group $%
\frac{ SO(5,1)}{O(4,1)}$and $f_{ijkl}^{pq}$ of $\frac{SO(6)}{O(5)}$.

Using 
\begin{eqnarray}
x_{+} &=&t+x\; ,\quad x_{-}=t-x\; ,\quad J_{\pm \ ab}=J_{0\ ab}\pm J_{1\
ab}\; ,  \notag \\
J_{\pm \ ab} &=&J_{\pm \ ab}(x_{\pm })\quad ,
\end{eqnarray}%
we can deduce from the equal-time commutation relations the commutation
rules for any space-time point, 
\begin{eqnarray}
\lbrack J_{\pm \ ab}(x_{\pm }),J_{\pm \ cd}(y_{\pm })]
&=&2if_{abcd}^{ef}J_{\pm \ ef}(x_{\pm })\delta (x_{\pm }-y_{\pm })  \notag \\
&&+2iC_{1}\delta _{ac}\delta _{bd}\delta ^{\prime }(x_{\pm }-y_{\pm }) 
\notag \\
&&+2iC_{2}\delta _{ad}\delta _{bc}\delta ^{\prime }(x_{\pm }-y_{\pm }) \; , 
\notag \\
\lbrack J_{+\ ab}(x_{+}),J_{-\ cd}(y_{-})] &=&0 \quad .  \notag \\
&&
\end{eqnarray}%
We can now decompose also the currents $J_{\pm \ ab}$, into creation and
annihilation parts, each one of massless excitations. We have%
\begin{equation}
J_{\pm \ ab}(x_{\pm })=J_{\pm \ ab}^{(+)}(x_{\pm })+J_{\pm \
ab}^{(-)}(x_{\pm })
\end{equation}%
where $(+)$ is the creation part and $(-)$ the annihilation part. Note that
here two creation or two annihilation operators of different $\frac{SO(5,1)}{%
O(4,1)}$ indices do not commute.

One finds also 
\begin{eqnarray}
&&\lbrack J_{\pm \ ab}(x_{\pm }),\psi _{ic}(x_{+}^{\prime },x_{-}^{\prime })]
\notag \\
&=&-\sigma \delta _{ab}(1\pm \delta \gamma _{5})\psi _{ic}(x_{+}^{\prime
},x_{-}^{\prime })\delta (x_{\pm }-x_{\pm }^{\prime })
\end{eqnarray}%
where, due to Jacobi identity $\sigma =1$ and $\delta ^{2}=1$. A similar
construction with the current $J_{\mu ij}$ can be trivially obtained.

Correlation functions are now immediately obtained from the methods of
two-dimensional conformally invariant Quantum Field Theory \cite{aar}.

The by now rather expected results displayed above mean that integrable
models can have a conformally invariant counterpart. The fact that in string
theory one needs conformal invariance as a building block forces us into the
above solution at least for the fermionic models in question.

The rather important unanswered question is about what happens in case of a
purely bosonic theory, or also, maybe even more important, to the model
defined on a graded manifold. In the last case, in view of the unbroken
supersymmetry, we are led to a conjecture concerning such sigma models,
namely we conjecture that such models have a conformal fix point where the
correlators are exactly solvable and present the previous symmetry.

\section*{Acknowledgments}

%(\textit{{Acknowledgments}}).
This work has been supported by {\small FAPESP} and {\small CNPq}, Brazil.

\end{document}